\documentclass[aps,pra,showpacs,groupedaddress,floatfix,nofootinbib]{revtex4}

\usepackage{graphicx}

\begin{document}

\title{PT-symmetry broken by point-group symmetry}
\author{Francisco M Fern\'andez} \email{fernande@quimica.unlp.edu.ar}
\author{Javier Garcia}

\affiliation{INIFTA (UNLP, CCT La Plata-CONICET), Divisi\'on
Qu\'imica Te\'orica, Blvd. 113 S/N,  Sucursal 4, Casilla de Correo
16, 1900 La Plata, Argentina}

\begin{abstract}
We discuss a PT-symmetric Hamiltonian with complex eigenvalues. It is based
on the dimensionless Schr\"{o}dinger equation for a particle in a square box
with the PT-symmetric potential $V(x,y)=iaxy$. Perturbation theory clearly
shows that some of the eigenvalues are complex for sufficiently small values
of $|a|$. Point-group symmetry proves useful to guess if some of the
eigenvalues may already be complex for all values of the coupling constant.
We confirm those conclusions by means of an accurate numerical calculation
based on the diagonalization method. On the other hand, the Schr\"odinger
equation with the potential $V(x,y)=iaxy^{2}$ exhibits real eigenvalues for
sufficiently small values of $|a|$. Point group symmetry suggests that
PT-symmetry may be broken in the former case and unbroken in the latter one.
\end{abstract}


\maketitle

\section{Introduction}

\label{sec:intro}

It was shown some time ago that some complex non-Hermitian
Hamiltonians may exhibit real eigenvalues\cite{CGM80, A95}. The
conjecture that such intriguing feature may be due to unbroken
PT-symmetry\cite{BB98} gave rise to a very active field of
research\cite{B07} (and references therein). The first studied
PT-symmetric models were mainly one-dimensional anharmonic
oscillators\cite{BB98,B07,FGZ98,FGRZ99} and lately the focus
shifted towards multidimensional
problems\cite{BDMS01,NA02,N02,N05,BTZ06,W09,BW12,HV13}. Among the
most widely studied multidimensional PT-symmetric models we
mention the complex versions of the Barbanis\cite
{BDMS01,NA02,N05,BTZ06,W09,BW12,HV13} and
H\'{e}non-Heiles\cite{BDMS01,W09} Hamiltonians. Several methods
have been applied to the calculation of their spectra: the
diagonalization method\cite{BDMS01,NA02,N02,N05,W09,BW12},
perturbation theory\cite{BDMS01,N02,N05,W09}, classical and
semiclassical approaches\cite{BDMS01,NA02}, among
others\cite{W09,HV13}. Typically, those models depend on a
potential parameter $g$ so that the Hamiltonian is Hermitian when
$g=0$ and non-Hermitian when $g\neq 0$. Bender and Weir\cite
{BW12} conjectured that some of those models may exhibit phase
transitions so that their spectra are real for sufficiently small
values of $|g|$. Such phase transitions appear to be a high-energy
phenomenon.

Multidimensional oscillators exhibit point-group symmetry (PGS)\cite
{PE81a,PE81b}. As far as we know such a property has not been taken into
consideration in those earlier studies of the PT-symmetric models, except
for the occasional parity in one of the variables. It is more than likely
that PGS may be relevant to the study of the spectra of multidimensional
PT-symmetric Hamiltonians. This paper is expected to be a useful
contribution in that direction.

The research on non-Hermitian Hamiltonians has been mainly focussed on
finding models with real spectrum. It is our purpose to show an example of
PT-symmetric Hamiltonian with complex eigenvalues; a Hamiltonian with the
phase transition at the Hermitian limit $g=0$. We will also show that PGS
provides a simple and clear explanation of why the eigenvalues of such model
are complex and not real as the other problems discussed so far.

In section~\ref{sec:model} we consider the dimensionless Schr\"{o}dinger
equation for a particle in a square box with the potential $iaxy$ that is
obviously PT-symmetric. In section~\ref{sec:PT} we show that perturbation
theory predicts that some of the eigenvalues are complex for sufficiently
small values of $|a|$. In section~\ref{sec:PGS} we analyze the
eigenfunctions of the unperturbed and perturbed Hamiltonians from the point
of view of PGS and show why some eigenvalues are expected to be complex. In
section~\ref{sec:DM} we obtain the eigenvalues and eigenfunctions accurately
by means of the diagonalization method and confirm the conclusions of the
preceding sections. In section~\ref{sec:Model_2} we consider the particle in
a square box with the potential $iaxy^{2}$ that resembles part of the
potential of the PT-symmetric version of the Barbanis Hamiltonian. In this
case PGS shows that PT symmetry may not be broken for sufficiently small
values of $|a|$. This conclusion is confirmed by the diagonalization method.
Finally, in section~\ref{sec:conclusions} we summarize the main results of
the paper, draw conclusions and put forward a somewhat general recipe for
the appearance of complex eigenvalues in a given multidimensional
non-Hermitian Hamiltonian.

\section{Box model with $C_{2v}$ point-group symmetry}

\label{sec:model}

We first consider the Schr\"{o}dinger equation $H\psi =E\psi $ with the
dimensionless Hamiltonian operator
\begin{equation}
H=p_{x}^{2}+p_{y}^{2}+gxy,  \label{eq:H_model}
\end{equation}
and the boundary conditions
\begin{equation}
\psi (\pm 1,y)=0,\;\psi (x,\pm 1)=0 .  \label{eq:BC}
\end{equation}
This Hamiltonian is Hermitian when $g$ is real and PT-symmetric when $g$ is
imaginary. In fact, when $g=ia$, $a$ real, the Hamiltonian is invariant
under two antiunitary transformations\cite{W60}
\begin{equation}
A_{x}HA_{x}=H,\;A_{y}HA_{y}=H  \label{eq:AHA=H}
\end{equation}
generated by $A_{x}=P_{x}T$ and $A_{y}=P_{y}T$, where $T$ is the
time-reversal operator\cite{P65} and $P_{x}$ and $P_{y}$ are the parity
transformations
\begin{eqnarray}
P_{x} &:&(x,y,p_{x},p_{y})\rightarrow (-x,y,-p_{x},p_{y}),  \nonumber \\
P_{y} &:&(x,y,p_{x},p_{y})\rightarrow (x,-y,p_{x},-p_{y}).  \label{eq:Px_Py}
\end{eqnarray}

It follows from equation (\ref{eq:AHA=H}) that
\begin{equation}
HA_{x}\psi =A_{x}H\psi =A_{x}E\psi =E^{*}A_{x}\psi .  \label{eq:HAxpsi}
\end{equation}
That is to say, if $\psi $ is eigenfunction of $H$ with eigenvalue $E$ then $%
A_{x}\psi $ is eigenfunction with eigenvalue $E^{*}$. Obviously, the same
conclusion applies to $A_{y}\psi $. When PT symmetry is unbroken
\begin{equation}
A_{x}\psi =\lambda \psi ,\;|\lambda |=1 ,  \label{eq:unbrok_sym}
\end{equation}
the corresponding eigenvalue is real\cite{B07}. In a recent paper we have
shown that the eigenvalue may be real even when this condition is manifestly
violated\cite{FG13}. Later on we will discuss this point in more detail. All
the Hamiltonians studied previously exhibit unbroken PT symmetry for
sufficiently small values of $|g|$\cite
{BDMS01,NA02,N02,N05,BTZ06,W09,BW12,HV13}. In what follows we show that the
model depicted above behaves in a quite different way.

\section{Perturbation theory}

\label{sec:PT}

When $g=0$ the eigenvalues and eigenfunctions of the simple model described
in the preceding section are those of the particle in a square box
\begin{eqnarray}
E_{mn}^{(0)} &=&\frac{\left( m^{2}+n^{2}\right) \pi ^{2}}{4}%
,\;m,n=1,2,\ldots ,  \nonumber \\
\psi _{mn}^{(0)}(x,y) &=&\varphi _{mn}(x,y)=\sin \left( \frac{m\pi (x+1)}{2}%
\right) \sin \left( \frac{n\pi (y+1)}{2}\right) ,  \label{eq:E(0),psi(0)}
\end{eqnarray}
and we appreciate that the eigenfunctions with $m\neq n$ are
two-fold degenerate. There are accidental degeneracies that occur
when $m_{1}^{2}+n_{1}^{2}=m_{2}^{2}+n_{2}^{2}$ but they are not
relevant for the present discussion. For example, the three
eigenfunctions $\psi _{7\,1}^{(0)} $, $\psi _{1\,7}^{(0)}$ and
$\psi _{5\,5}^{(0)}$ share the same eigenvalue but only the first
two ones are consequence of the symmetry of the problem.

By means of perturbation theory it is quite easy to prove that some of the
eigenvalues are complex for sufficiently small values of $|a|$. The
perturbation correction of first order $E_{mn}^{(1)}$ vanishes when $n=m+2j$%
, $j=0,1,\ldots $ but it is nonzero if $n=m+2j+1$:

\begin{eqnarray}
E_{mn}^{(1)+} &=&\frac{256m^{2}\left( 2j+m+1\right) ^{2}}{\pi ^{4}\left(
2j+1\right) ^{4}\left( 2j+2m+1\right) ^{4}} ,  \nonumber \\
E_{mn}^{(1)-} &=&-\frac{256m^{2}\left( 2j+m+1\right) ^{2}}{\pi ^{4}\left(
2j+1\right) ^{4}\left( 2j+2m+1\right) ^{4}} .  \label{eq:Emn(1)}
\end{eqnarray}
It is clear that for sufficiently small values of $|a|$ these levels behave
approximately as linear functions of $g=ia$. In other words, the phase
transition takes place at the Hermitian limit $a=0$. This result is
different from that for the PT-symmetric oscillators studied so far that
exhibit a vanishing perturbation correction of first order\cite{BDMS01,W09}.

\section{Point-group symmetry}

\label{sec:PGS}

We can understand the occurrence of complex eigenvalues more clearly from
the point of view of PGS. Since the model is two-dimensional its behaviour
with respect to the coordinate $z$ is irrelevant and, consequently, the
choice of the point group is not unique. For the description of the
unperturbed model $g=0$ we choose the point group $C_{4v}$ with symmetry
operations
\begin{eqnarray}
E &:&(x,y)\rightarrow (x,y),  \nonumber \\
C_{4} &:&(x,y)\rightarrow (y,-x),  \nonumber \\
C_{4}^{3} &:&(x,y)\rightarrow (-y,x),  \nonumber \\
C_{2} &:&(x,y)\rightarrow (-x,-y),  \nonumber \\
\sigma _{v1} &:&(x,y)\rightarrow (y,x),  \nonumber \\
\sigma _{v2} &:&(x,y)\rightarrow (-y,-x),  \nonumber \\
\sigma _{d1} &:&(x,y)\rightarrow (x,-y),  \nonumber \\
\sigma _{d2} &:&(x,y)\rightarrow (-x,y),  \label{eq:C4v_op}
\end{eqnarray}
where $C_{n}^{k}$ is a rotation by an angle $2\pi k/n$ around an axis
perpendicular to the center of the square box ($C_{4}^{2}=C_{2}$) and $%
\sigma _{v}$ and $\sigma _{d}$ are vertical reflection
planes\cite{C90,T64}. For simplicity we omit the transformation of
the momenta when it is similar to that of the coordinates. The
eigenfunctions form bases for the irreducible representations
$\{A_{1},A_{2},B_{1},B_{2},E\}$ as indicated below
\begin{eqnarray}
A_{1} &:&\{\varphi _{2m-1\,2m-1}\},\;\{\varphi _{2m-1\,2n-1}^{+}\},
\nonumber \\
A_{2} &:&\{\varphi _{2m\,\,2n}^{-}\},  \nonumber \\
B_{1} &:&\{\varphi _{2m\,2m}\},\;\{\varphi _{2m\,2n}^{+}\},  \nonumber \\
B_{2} &:&\{\varphi _{2m-1\,2n-1}^{-}\},  \nonumber \\
E &:&\{\varphi _{2m-1\,2n},\varphi _{2n\,2m-1}\},  \nonumber \\
\varphi _{m\,n}^{\pm } &=&\frac{1}{\sqrt{2}}\left( \varphi _{m\,n}\pm
\varphi _{n\,m}\right)  \nonumber \\
m,n &=&1,2,\ldots .  \label{eq:unpert_eigenf}
\end{eqnarray}
As expected some pairs of two-fold degenerate eigenfunctions form bases for
the irreducible representation $E$. In addition to it, pairs of
eigenfunctions with symmetry $A_{1}$ and $B_{2}$ ($\varphi _{2m-1\,2n-1}^{+}$%
, $\varphi _{2m-1\,2n-1}^{-}$) as well as $A_{2}$ and $B_{1}$ ($\varphi
_{2m\,\,2n}^{-}$, $\varphi _{2m\,2n}^{+}$) are also degenerate.

When $g\neq 0$ a suitable point group is $C_{2v}$ with symmetry operations $%
\{E,C_{2},\sigma _{v1},\sigma _{v2}\}$ and irreducible representations $%
\{A_{1},A_{2},B_{1},B_{2}\}$. The eigenfunctions are linear combinations of
the form
\begin{eqnarray}
\psi ^{A_{1}} &=&\sum_{m}\sum_{n}\left( a_{mn}^{A_{1}}\varphi
_{2m-1\,2m-1}+b_{mn}^{A_{1}}\varphi _{2m\,2m}+c_{mn}\varphi
_{2m-1\,2n-1}^{+}+d_{mn}\varphi _{2m\,2n}^{+}\right)  \nonumber \\
\psi ^{A_{2}} &=&\sum_{m}\sum_{n}\left( a_{mn}^{A_{2}}\varphi
_{2m-1\,2n-1}^{-}+b_{mn}^{A_{2}}\varphi _{2m\,2n}^{-}\right)  \nonumber \\
\psi ^{B_{1}} &=&\sum_{m}\sum_{n}a_{mn}^{B_{1}}\varphi _{2m-1\,2n}^{+}
\nonumber \\
\psi ^{B_{2}} &=&\sum_{m}\sum_{n}a_{mn}^{B_{2}}\varphi _{2m-1\,2n}^{-}
\label{eq:pert_eigenf}
\end{eqnarray}
It is clear that the perturbation removes the degeneracy in such a way that
the two-fold degenerate unperturbed eigenfunctions $E$ become the perturbed
eigenfunctions of symmetry $B_{1}$ and $B_{2}$. As a result, every
eigenvalue $E_{B_{1}}$ is the complex conjugate of an eigenvalue $E_{B_{2}}$
($E_{B_{2}}=E_{B_{1}}^{*}$). As shown in the preceding section, the
degeneracy of these levels is removed at first order of perturbation theory
and it is not difficult to verify that the pair of integrals $\left\langle
\varphi _{2m\,2n-1}^{\pm }\right| xy\left| \varphi _{2m\,2n-1}^{\pm
}\right\rangle $ give us exactly the perturbation corrections in equation (%
\ref{eq:Emn(1)}). On the other hand, the degenerate unperturbed
eigenfunctions of symmetry $A_{1}$, $A_{2}$, $B_{1}$ and $B_{2}$ become the
perturbed eigenfunctions of symmetry $A_{1}$ and $A_{2}$. In this case the
perturbation correction of first order vanishes and the degeneracy is
removed at least at second order. If, as in the case of the models studied
earlier by other authors, all the perturbation corrections of odd order
vanish\cite{BDMS01,W09}, then we may expect real eigenvalues for
sufficiently small values of $|a|$.

PGS gives us a clear description of the occurrence of complex
eigenvalues. If we take into account equation (\ref{eq:HAxpsi})
and that $A_{x}\varphi _{2m-1\,2n}^{+}=-\varphi _{2m-1\,2n}^{-}$
then we realize that $A_{x}\psi ^{B_{1}}=\lambda _{B_{1}B_{2}}\psi
^{B_{2}}$. We appreciate that PT symmetry is broken for all
$|g|\neq 0$ and that $E_{B_{2}}=E_{B_{1}}^{*}$ as mentioned above.
However, in principle it may be possible that both eigenvalues
were real and degenerate as in the case of the rigid rotor studied
in an earlier paper\cite{FG13}. In the present case we know that
they are complex as shown in section~\ref{sec:PT}. If we apply the
same reasoning to the eigenfunctions of symmetry $A_{1}$ and
$A_{2} $ we realize that PT-symmetry may not be broken for them
because $A_{x}\psi ^{A_{1}}=\lambda _{A_{1}}\psi ^{A_{1}}$ and
$A_{x}\psi ^{A_{2}}=\lambda _{A_{2}}\psi ^{A_{2}}$ (where
$|\lambda |=1$) that follows from the fact that the
symmetry-adapted basis functions are invariant or merely change
sign under this antiunitary operation (and also under $A_{y}$).

\section{Diagonalization method}

\label{sec:DM}

We can obtain sufficiently accurate eigenvalues and eigenfunctions of the
box model by means of the diagonalization method. Diagonalization of the
Hamiltonian matrix $\mathbf{H}$ in the basis set $\{\varphi _{mn}\}$ gives
us the lowest eigenvalues of the Hamiltonian operator as well as the
coefficients of the expansion of the eigenfunctions in the basis set
\begin{equation}
\psi =\sum_{m}\sum_{n}a_{mn}\varphi _{mn}.
\end{equation}
Alternatively, we can diagonalize Hamiltonian matrices $\mathbf{H}^{S}$ for
each of the irreducible representations $S=A_{1},A_{2},B_{1},B_{2}$ of the
point group $C_{2v}$ and thus obtain the corresponding sets of
eigenfunctions (\ref{eq:pert_eigenf}) separately. In this case the dimension
of the resulting secular equations is noticeably smaller.

It is well known that the coefficients of the characteristic polynomial
generated by the full matrix $\mathbf{H}$ are real\cite{F13}. The
coefficients of the characteristic polynomials generated by $\mathbf{H}%
^{A_{1}}$ and $\mathbf{H}^{A_{2}}$ are polynomial functions of $g^{2}$ and
therefore real. On the other hand, the coefficients of the characteristic
polynomials generated by the matrices $\mathbf{H}^{B_{1}}$ and $\mathbf{H}%
^{B_{2}}$ are polynomial functions of $g$ and therefore complex.

\begin{figure}[tbp]
~\bigskip\bigskip
\par
\begin{center}
\includegraphics[width=6cm]{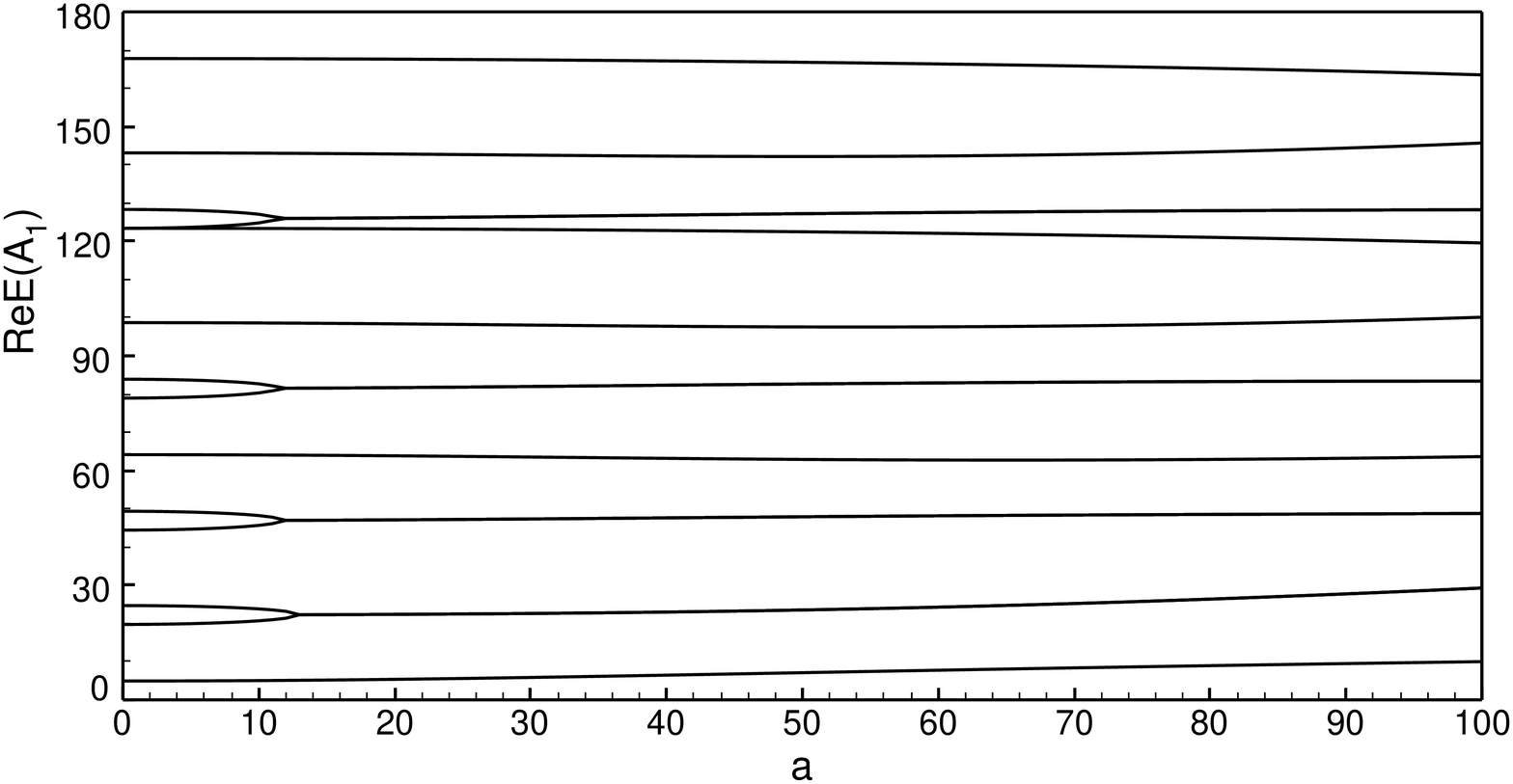} \includegraphics[width=6cm]{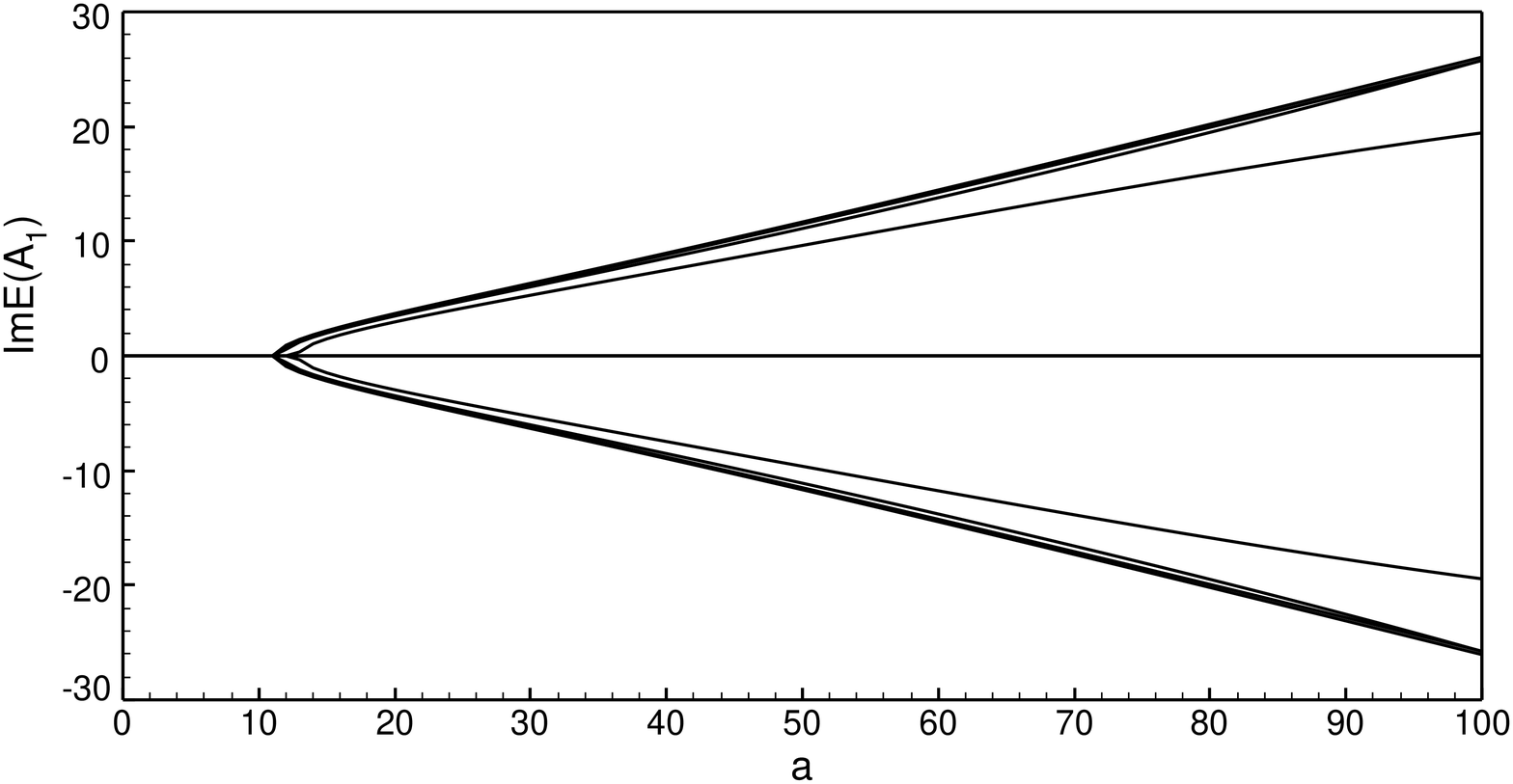}
\includegraphics[width=6cm]{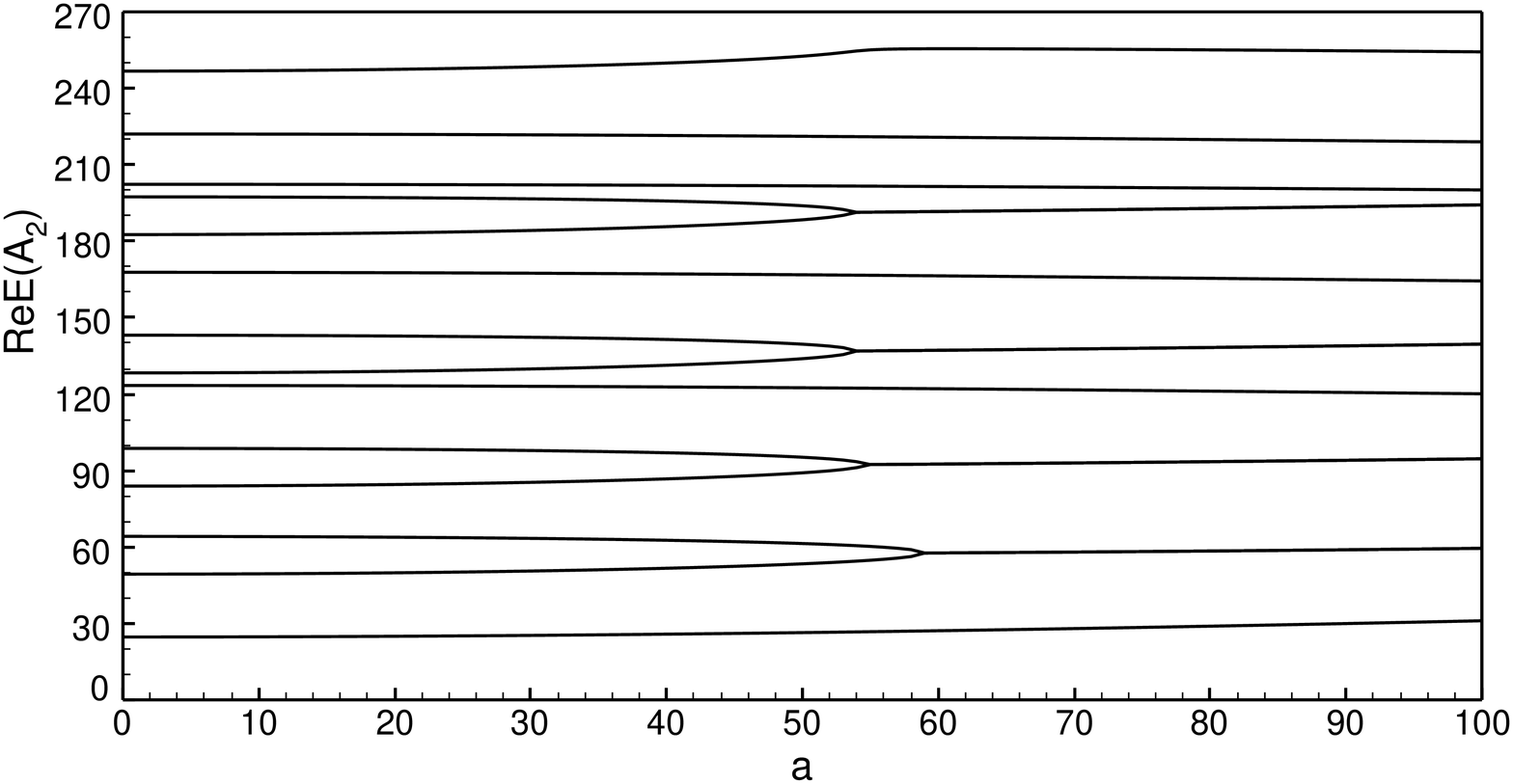} \includegraphics[width=6cm]{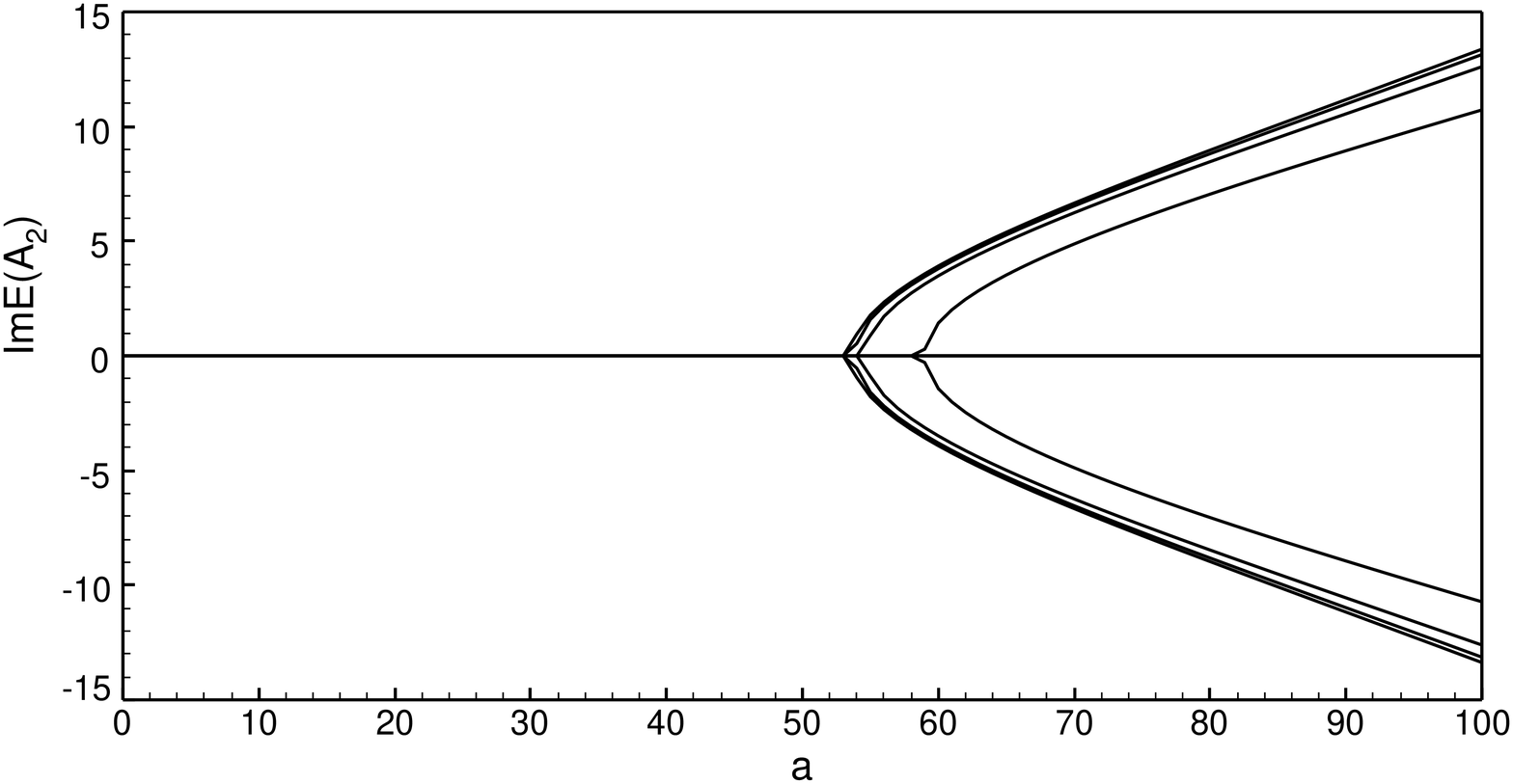}
\includegraphics[width=6cm]{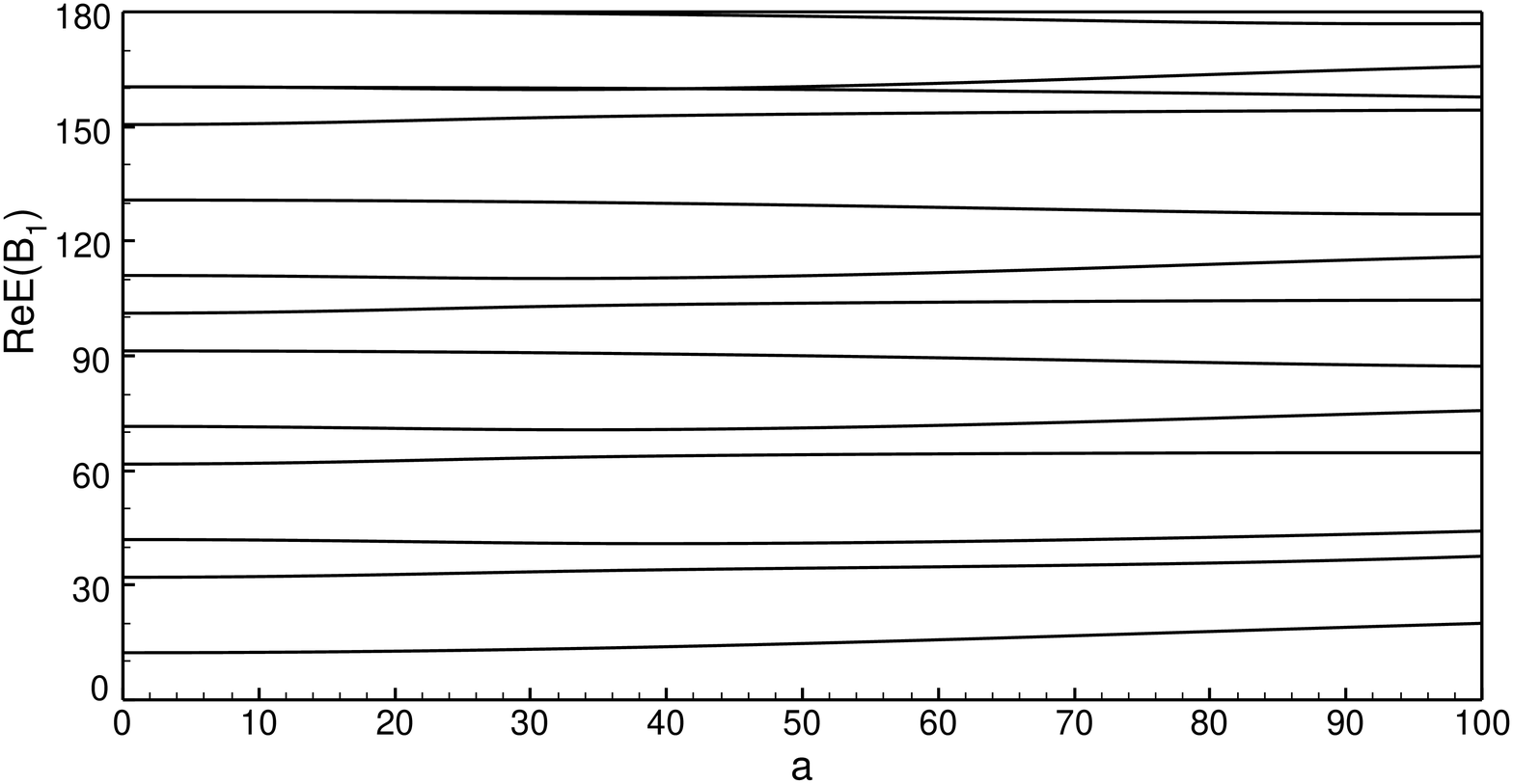} \includegraphics[width=6cm]{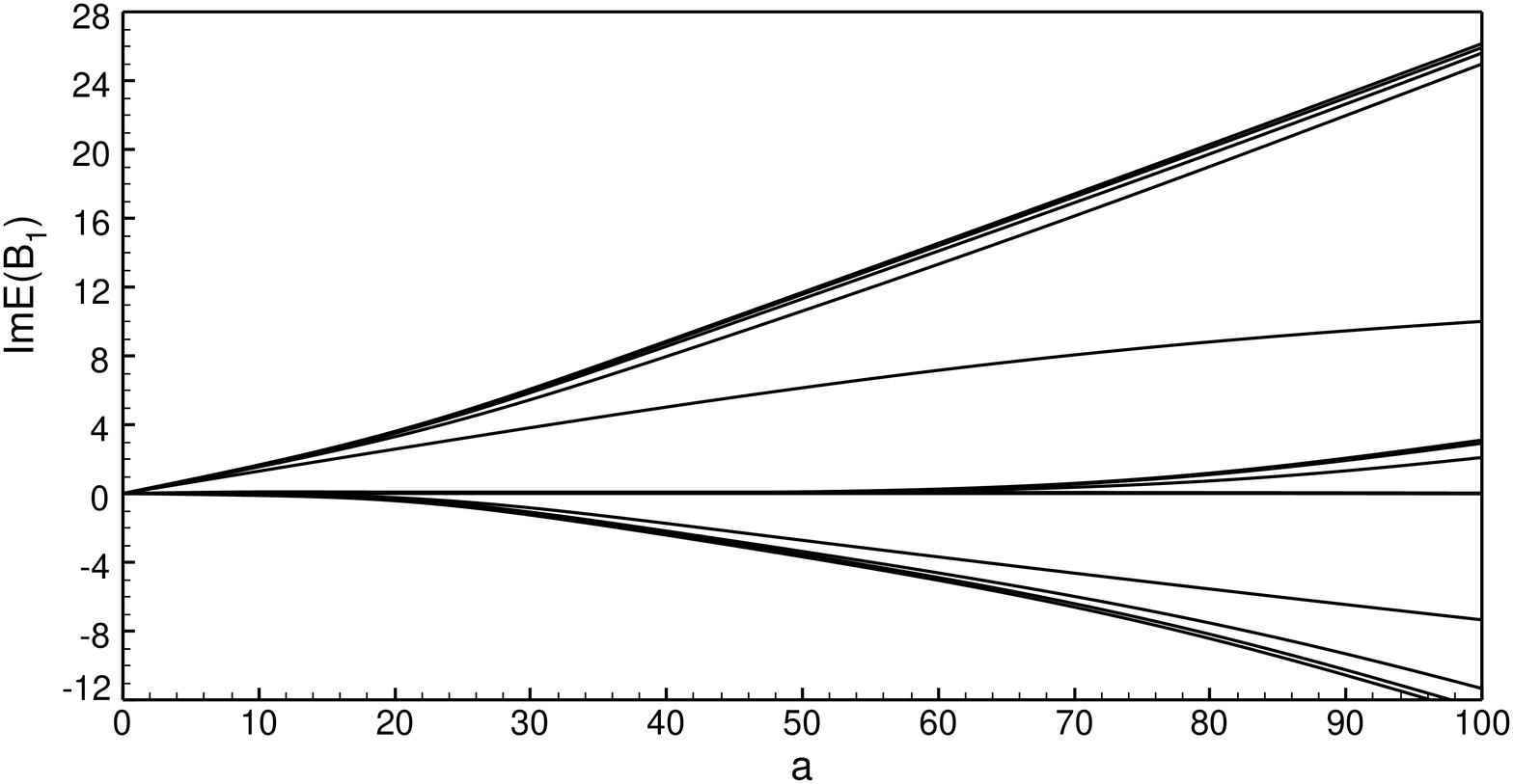}
\bigskip
\end{center}
\caption{First eigenvalues of the model (\ref{eq:H_model}) }
\label{fig:Emn}
\end{figure}

Figure~\ref{fig:Emn} shows the real and imaginary parts of the first
eigenvalues of the Hamiltonian (\ref{eq:H_model}) for a wide range of values
of $a$. The eigenvalues for symmetry $A_{1}$ and $A_{2}$ are real for
sufficiently small values of $a$. Some pairs of them coalesce at critical
values $a_{c}$ of the coupling constant and emerge as pairs of complex
numbers for $a>a_{c}$. This occurrence of exceptional points is similar to
that already found for other two-dimensional models\cite
{BDMS01,NA02,N02,N05,BTZ06,W09,BW12,HV13}. On the other hand, the
eigenvalues for symmetry $B_{1}$ and $B_{2}$ are complex for all values of $%
a $. This kind of eigenvalues does not appear in those non-Hermitian
Hamiltonians studied earlier.\ We say that the PT-symmetric Hamiltonian (\ref
{eq:H_model}) exhibits a PT phase transition at the trivial Hermitian limit.

\section{Box model with $C_{2}$ point-group symmetry}

\label{sec:Model_2}

In order to illustrate the difference between present PT-symmetric model and
those studied earlier, in this section we choose the particle in a square
box with the interaction potential
\begin{equation}
V(x,y)=gxy^{2}  \label{eq:V_M_2}
\end{equation}
that resembles the one in the Barbanis Hamiltonian\cite
{BDMS01,NA02,N05,BTZ06,W09,BW12,HV13}. In this case we may choose the point
group $C_{2}$ with symmetry operations $\{E,C_{2}\}$, where $%
C_{2}:(x,y)\rightarrow (x,-y)$. The bases for the irreducible
representations $\{A,B\}$ are $\{\varphi _{m\,2n-1}\}$ and $\{\varphi
_{m\,2n}\}$, respectively. The antiunitary operator $A=P_{x}T$, where $%
P_{x}:(x,y,p_{x},p_{y})\rightarrow (-x,y,-p_{x},p_{y})$, leaves the
Hamiltonian invariant when $g=ia,a$ real. It follows from $A\varphi
_{m\,n}=(-1)^{m+1}\varphi _{m\,n}$ and equation (\ref{eq:HAxpsi}) that it is
possible that $A\psi ^{A}=\lambda _{A}\psi ^{A}$ and $A\psi ^{B}=\lambda
_{B}\psi ^{B}$; that is to say symmetry may be unbroken and the eigenvalues
may be real for sufficiently small values of $|a|$. This observation is
consistent with the fact that the perturbation correction of first order
vanishes for all the states which suggests that the perturbation expansion
exhibits only even powers of $g$ as in the case of the Barbanis Hamiltonian%
\cite{BDMS01,W09}.

\begin{figure}[tbp]
~\bigskip\bigskip
\par
\begin{center}
\includegraphics[width=6cm]{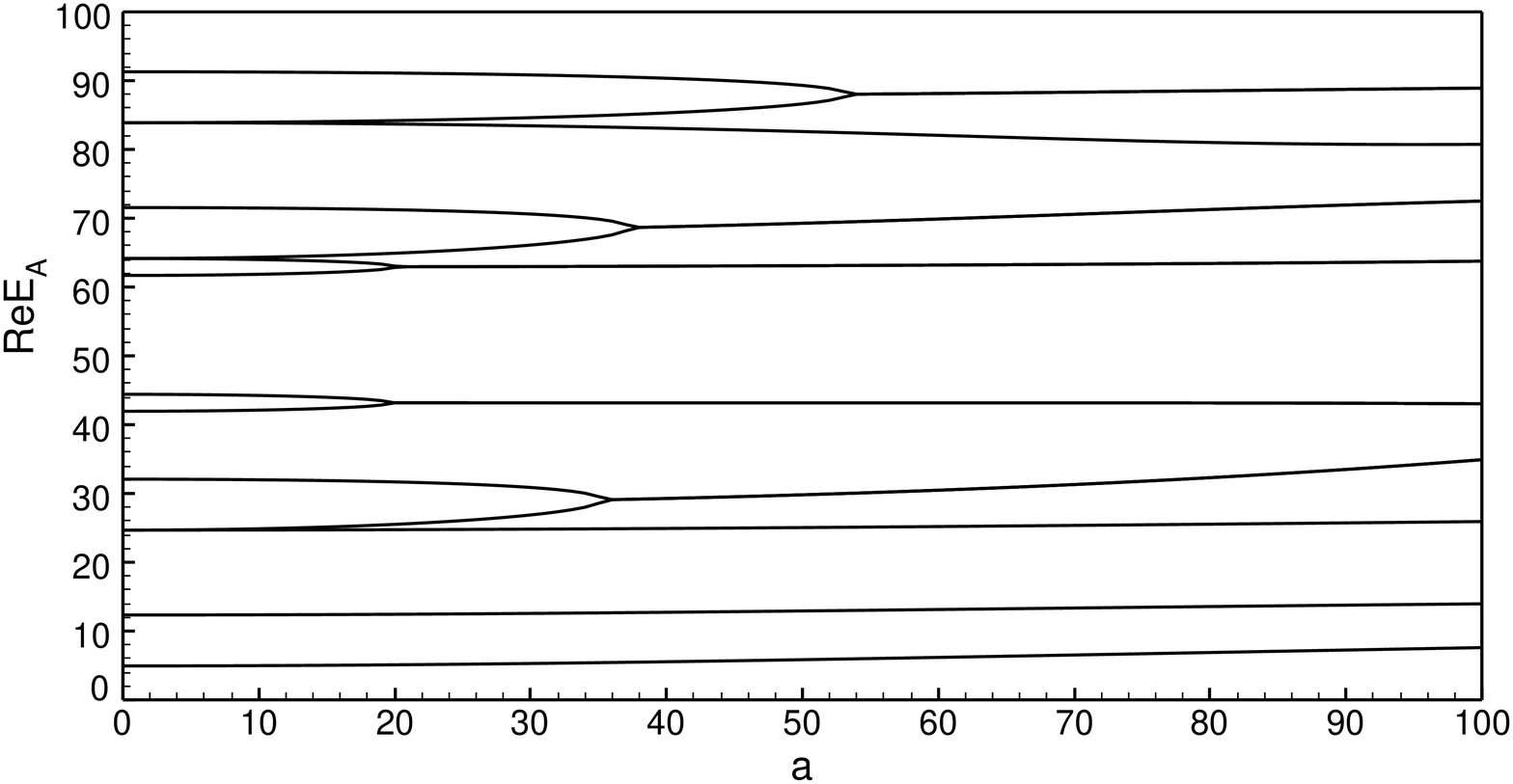} \includegraphics[width=6cm]{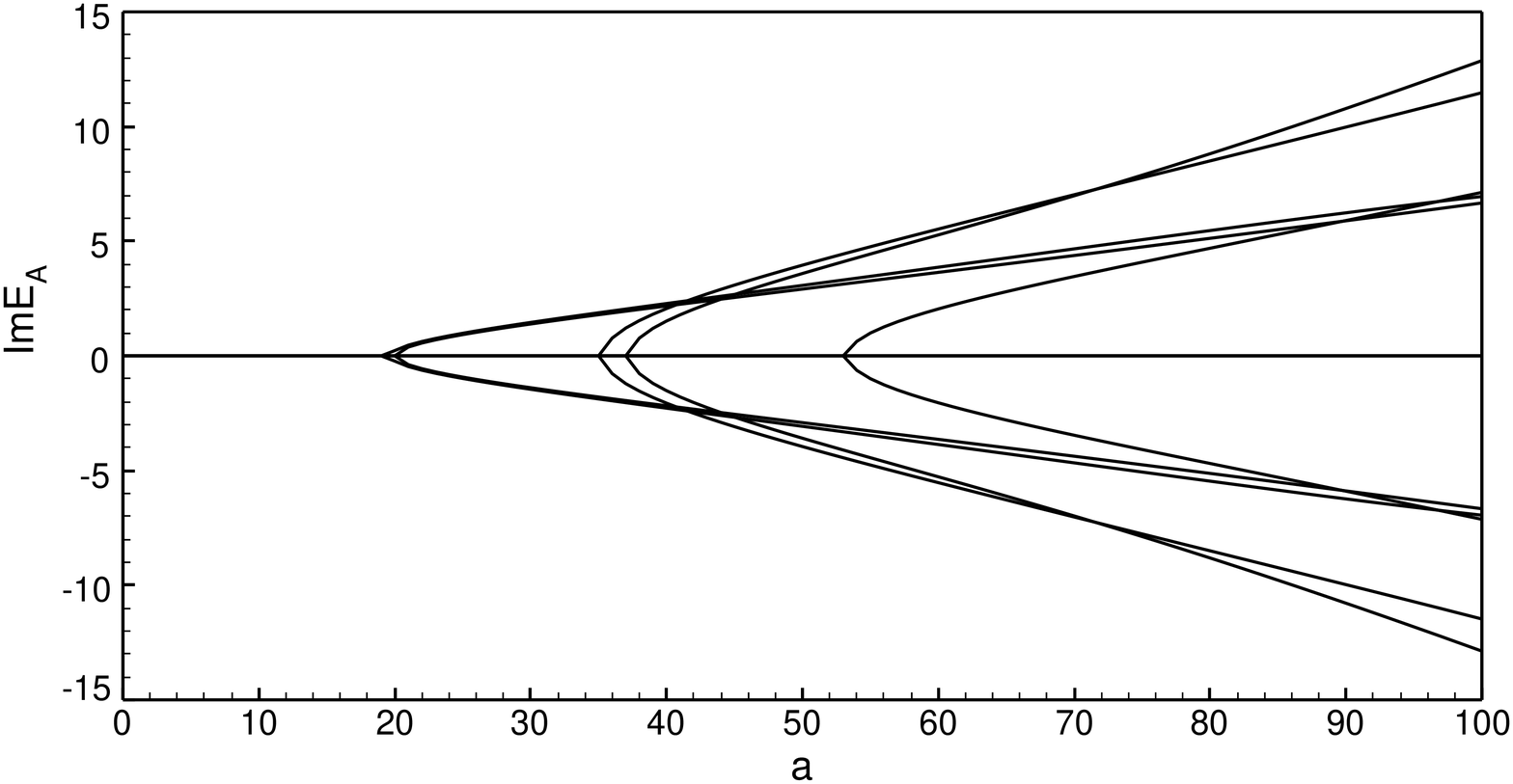} %
\includegraphics[width=6cm]{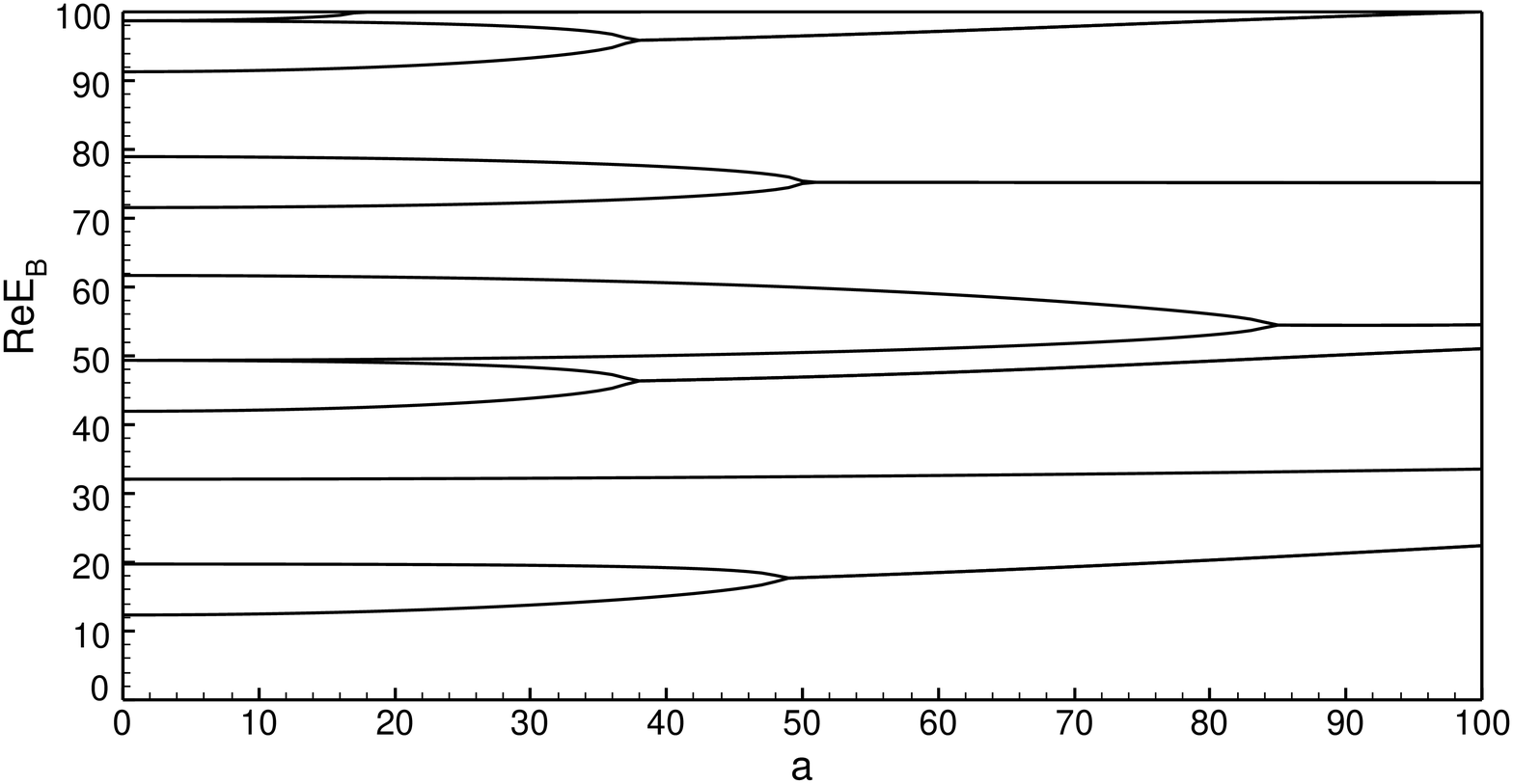} \includegraphics[width=6cm]{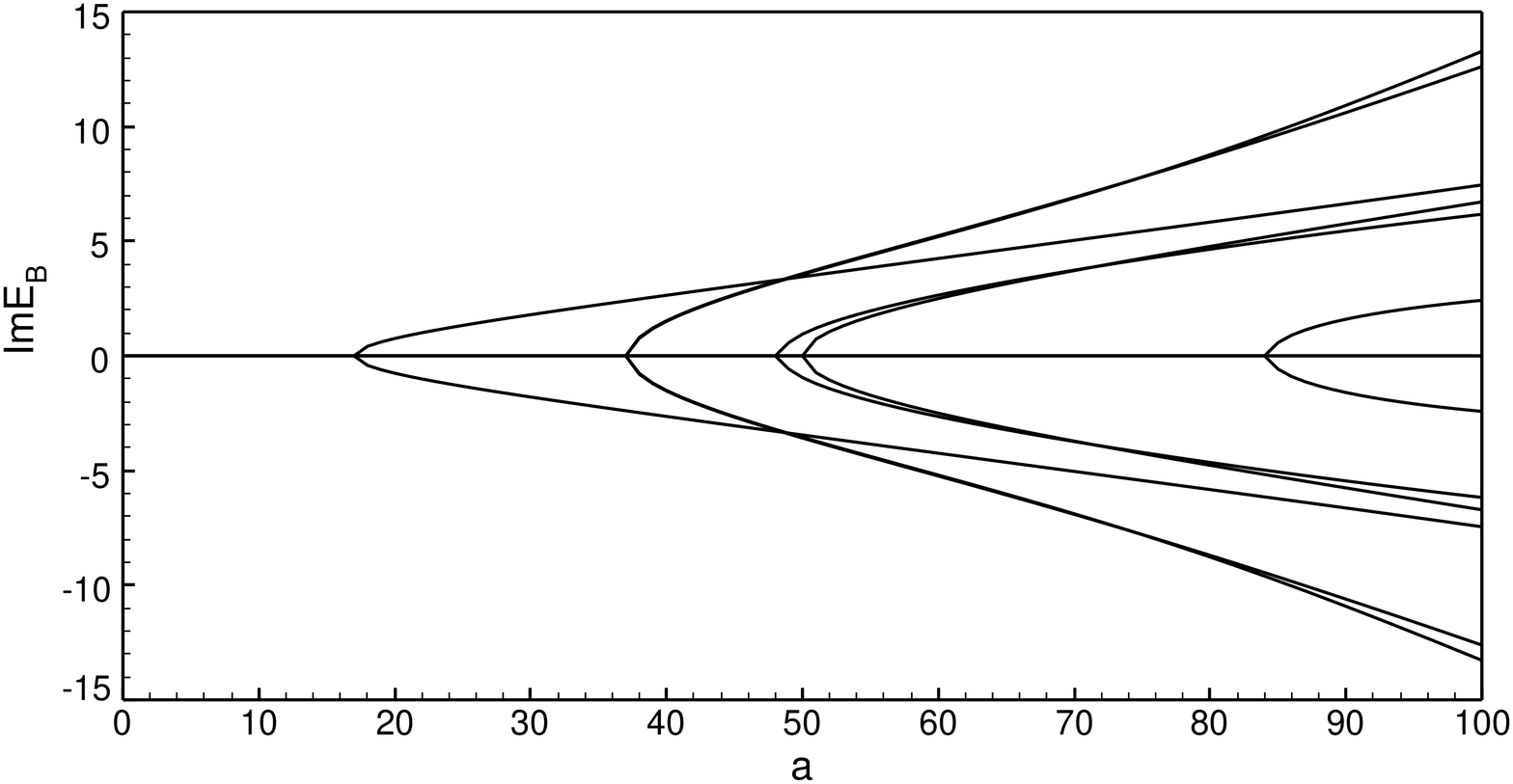}
\bigskip
\end{center}
\caption{First eigenvalues of the model (\ref{eq:V_M_2}) }
\label{fig:Emn_C2}
\end{figure}

The results of the diagonalization method are shown in figure~\ref
{fig:Emn_C2}. All the eigenvalues of both types of symmetry appear
to be real for sufficiently small values of $|a|$ as expected from
the argument based on PGS. Both sets of eigenvalues exhibit
exceptional points where a pair of eigenvalues coalesce and emerge
as complex conjugate numbers. The main features of the spectrum of
this model resemble those described in earlier problems.

\section{Conclusions}

\label{sec:conclusions}

In this paper we have discussed two non-Hermitian Hamiltonians with
completely different spectra. Both are representative of a wider class of
non-Hermitian Hamiltonians that depend on a parameter $g$ in such a way that
they are Hermitian when $g=0$ and PT-symmetric for $g$ nonzero and imaginary
(say, $g=ia$, $a$ real). The $C_{2}$ model is similar to those studied
earlier that were chosen in such a way that all the eigenvalues are real
when $0<a<a_{c}$\cite{BDMS01,NA02,N02,N05,BTZ06,W09,BW12,HV13}. It is said
that they exhibit a PT phase transition at $a=a_{c}$ that was conjectured to
be a high-energy phenomenon\cite{BW12} (at least for those examples).
Consistent with real eigenvalues is the fact that their perturbation series
exhibit only even powers of $g$\cite{BDMS01,N02,N05,W09}.

The $C_{2v}$ model is the main goal of this paper because it appears to
exhibit complex eigenvalues for all values of $a\neq 0$. Therefore, in this
case the PT phase transition takes place at the trivial Hermitian limit. The
reason for this behaviour is that the degeneracy of the unperturbed
Hamiltonian is broken at first order of perturbation theory so that the
eigenvalues are almost linear functions of $g=ia$ for sufficiently small
values of $|a|$.

It has been our purpose to show that PGS is quite useful for the
study of the spectra of such non-Hermitian Hamiltonians. The
analysis of the eigenfunctions from such point of view clearly
shows that PT symmetry is broken for those that are bases for some
irreducible representations ($B_{1}$ and $B_{2}$ in the present
case). We may formulate the main ideas in a somewhat more general
way. In general, the eigenfunctions are of the form
\begin{equation}
\psi ^{S}=\sum_{j}c_{j}^{S}\varphi _{j}^{S}
\end{equation}
where $S$ is an irreducible representation of the point group for
the model. If $A\varphi _{j}^{S}=\lambda _{j}^{S}\varphi
_{j}^{S}$, were $A$ is an antiunitary operation that leaves the
Hamiltonian invariant, then it is possible that $A\psi
^{S}=\lambda _{S}\psi ^{S}$ (unbroken PT symmetry) and the
corresponding eigenvalues are real. All the models studied before
exhibit this
property\cite{BDMS01,NA02,N02,N05,BTZ06,W09,BW12,HV13}. The
eigenfunctions of symmetry $A_{1}$ and $A_{2}$ of present $C_{2v}$
model also behave in this way. The situation is quite different in
the case of the eigenfunctions of symmetry $B_{1}$ and $B_{2}$
that we may generalize it in the following way: when $A\varphi
_{j}^{S}=\lambda _{k}^{SS^{\prime }}\varphi _{k}^{S^{\prime }}$
where $S^{\prime }\neq S$ then PT symmetry is
broken $A\psi ^{S}=\lambda _{SS^{\prime }}\psi ^{S^{\prime }}$ and $%
E_{S}=E_{S^{\prime }}^{*}$. However, this relationship is not a rigorous
proof that the eigenvalues are complex. In a recent paper we have shown that
the eigenvalues of a PT symmetric rigid rotor are real even when PT-symmetry
is broken as just indicated\cite{FG13}. In the case of present $C_{2v}$
model the eigenvalues are in fact complex and for this reason we have
decided to coin the term \textit{PT-symmetry broken by PGS}.

In closing, we want to put forward an additional argument. Let
$H_{R}$ and $H_{I}$ be the real and imaginary parts of the
non-Hermitian Hamiltonian $H$, and assume that $H_{R}$ is
invariant under the inversion operator
$\hat{\imath}:(\mathbf{x},\mathbf{p})\rightarrow
(-\mathbf{x},-\mathbf{p})$. Then the product $\varphi _{i}\varphi
_{j}$ of degenerate eigenfunctions of $H_{R}$ is invariant under
$\hat{\imath}$ as shown by the character tables of the symmetry
point groups\cite{C90,T64}. Under such conditions, if
$\hat{\imath}H_{I}\hat{\imath}=-H_{I}$ then $\left\langle \varphi
_{i}\right| H_{I}\left| \varphi _{j}\right\rangle =0$ and the
perturbation correction of first order vanishes. The non-Hermitian
Hamiltonian $H$ is invariant under the antiunitary operator
$A=\hat{\imath}T$ and it is possible that the spectrum be real.
If, on the other hand, $\hat{\imath}H_{I}\hat{\imath}=H_{I}$ then
some of the perturbation corrections of first order for the
degenerate states may not vanish and the corresponding eigenvalues
are expected to be complex.

In section~\ref{sec:PGS} $C_{2}$ plays the role of $\hat{\imath}$
and we appreciate that both $H_{R}=p_{x}^{2}+p_{y}^{2}$ and
$H_{I}=axy$ are invariant under $C_{2}$. For this reason there are
complex eigenvalues even though the non-Hermitian Hamiltonian is
invariant under $A_{x}$ and $A_{y}$ (it is not invariant under
$A$). On the other hand, the inversion operation changes the sign
of the potential in equation (\ref{eq:V_M_2}) and the perturbation
corrections of first order for all the states vanish. In this case
the non-Hermitian Hamiltonian is invariant under $A$ and exhibits
real spectrum for sufficiently small values of $a$.

It seems that the argument about unbroken PT symmetry\cite{B07}
applies to the antiunitary operator $A$ and not to other
antiunitary operators like $A_{x}$ and $A_{y}$.

\end{document}